# Links, Knots, and Knotted Labyrinths in Bistable Systems

Anatoly Malevanets and Raymond Kapral

*Chemical Physics Theory Group, Department of Chemistry, University of Toronto, Toronto, Ontario, Canada M5S 3H6*
(Received 5 March 1996)

The existence of stable links and knots is demonstrated in three-dimensional, bistable, chemical media. The reaction-diffusion medium segregates into regions of high and low concentration separated by sharp interfaces. The interfaces repel at short distances so that domains with various topologies are possible, depending on the initial conditions and system parameters. Front instabilities can give rise to knotted labyrinthine patterns. A lattice-gas model whose mean-field limit is the FitzHugh-Nagumo equation is described and implemented to carry out the simulations. [S0031-9007(96)00612-6]

Knots and links are encountered in many physical problems, as both concrete objects and mathematical constructs [1,2]. Some of the most direct examples are found in chemistry and biochemistry at the microscopic level; for example, knotted circular DNA has been observed [3], and topological arguments have been given to analyze its structure [4]. Linked and knotted filaments of phase singularities have been studied in excitable chemical media, and the implications of these structures for the nature of wave propagation processes have been investigated [5,6].

In this Letter we show that links and knots can arise in bistable chemical media when such systems segregate into two phases of low and high concentration, corresponding to the two stable states, separated by stationary, sharp interfaces. The evolution to stable knots with specific shapes allows one to use chemical dynamics to investigate the factors that determine stable knot geometries. The destabilization of these compact knots as system parameters are changed leads to disordered chemical patterns such as knotted labyrinths which may still possess the topological structure of knots.

Normally, in bistable chemical media with a nonconserved order parameter the more-stable phase will consume the less-stable phase until the medium becomes homogeneous. If the two phases have the same stability, one expects curvature-driven domain coarsening. However, it is known that bistable chemical (and other) systems composed of at least two relevant chemical species with different diffusion coefficients can display front repulsion at short distances leading to spatial structures such as the labyrinthine patterns observed in the iodide-ferrocaynide-sulfite system [7]. There have been extensive theoretical discussions of the instabilities giving rise to fronts and patterns of this type [8–11].

The FitzHugh-Nagumo (FHN) model,
$$u_t = -u^3 + u - v + D_u \nabla^2 u,$$
$$v_t = \epsilon(u - av + b) + D_v \nabla^2 v,$$
has been used to investigate the front dynamics in bistable chemical media in both the fast inhibitor limit ($\epsilon \gg 1$, $D_v/\epsilon \equiv \hat{D}_v$) [11] and the slow inhibitor limit ($\epsilon \ll 1$) [8,10] with $U$ the activator and $V$ the inhibitor.

The results reported here were obtained by utilizing a microscopic lattice-gas model for the dynamics whose mean field limit yields FHN kinetics. Consequently, we are able to verify that the chemical patterns we observe are robust with respect to fluctuations.

An underlying microscopic dynamics is not usually associated with the FHN model since its origin lies in the physiology of nerve impulse propagation [12]. However, it is possible to construct such a model using the following rules. The dynamics takes place on a three-dimensional (3D) simple cubic lattice. One may imagine two such lattices, $\mathcal{L}_u$ and $\mathcal{L}_v$, one for each of the $U$ and $V$ species. A node on $\mathcal{L}_i$ ($i = u, v$) may be occupied by $n_i$ particles, with $n_i \in \{0, 1, \ldots, m\}$. The reactions involve not only the $U$ and $V$ species themselves but also the number of "holes" or vacancies at a node. We denote such holes as $U^*$ or $V^*$ and their local number by $n_i^* = m - n_i$. Consider the following reaction scheme:

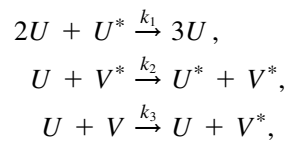

$$2U + U^* \xrightarrow{k_1} 3U,$$
$$U + V^* \xrightarrow{k_2} U^* + V^*,$$
$$U + V \xrightarrow{k_3} U + V^*,$$

plus a similar set of three equations with $U \leftrightarrow U^*$ and $V \leftrightarrow V^*$ characterized by rate constants $k_j^*$ ($j = 1, 2, 3$). Local reactions are carried out with probabilities proportional to the rate constants $k_j$ and $k_j^*$ and the product of the local species concentrations.

Diffusion was implemented in a standard fashion as a collection of random walkers. At each time step a collection of particles on a lattice is moved in a randomly chosen direction. These random walkers are not independent; however, analysis shows that correlations are small and can be safely neglected. The simulations reported here use a diffusion ratio $D_v/D_u = 4$, which is equivalent to using a step of unit length for walkers on lattice $\mathcal{L}_u$ and double that for lattice $\mathcal{L}_v$. A complete description of statistical properties of the model will be presented elsewhere.

In the mean-field limit, particles are binomially distributed on the lattice. In this case, one may show that the local reaction rule, averaged over the binomial distribution, yields the FHN rate law after a suitable scaling of the $u$ and $v$ variables. The constraints $k_i^* = k_i$ ($i = 2, 3$) are necessary to recover the linear coupling of the FHN model in the mean-field approximation. Since explicit relations between the $k_j$ and $k_j^*$ and the parameters $\epsilon$, $a$, and $b$ in the FHN equation exist, it is possible to study the lattice-gas model in various dynamical regimes corresponding to the deterministic FHN kinetics. Indeed, we have verified that the front bifurcation phenomena observed in our microscopic model occur at parameter values predicted from a stability analysis of the FHN equation. The simulations reported below were carried out on a CAM-8 machine [13] on lattices with $(256)^3$ nodes. The maximum occupancy of a node was restricted to $m = 7$ particles.

Links, knots, and knotted labyrinths are three-dimensional extensions with complex topologies of the localized objects analyzed in [8]. The two-dimensional disk-shaped objects can be considered as planar sections of filled tubes, which are the building elements of our patterns. In the fast inhibitor limit, $\epsilon \gg 1$, a nonlocal free energy functional exists [8,11], and the form of the repulsive front interaction has been determined from the front equations [11]. Our simulations were carried out in the slow inhibitor limit $\epsilon \ll 1$ where no free-energy functional exists. Front interactions in this limit have been considered in [8]. Domain fusion is prevented by the long-range nature of the $v$ field which leads to interactions between fronts on distance scales that are large compared to the $u$ front profile widths [8,10].

While one may expect this front repulsion to preserve the topological structure of the system during evolution, it is possible that curvature effects can become larger than the repulsive front interactions. The velocity normal to the front is given by $v_n \approx v_f + d\kappa$, where $v_f$ is the planar front velocity, $d$ is a constant determined by system parameters, and $\kappa$ is the mean curvature of the surface [8,10]. Our simulations show that the time scale for relaxation to a surface with an approximately constant mean curvature (the tubular domain) is well separated from that for the global dynamics of the tubular segments so that $v_r \approx 0$. Figure 1 shows the evolution of a triangle-shaped domain composed of tubular segments containing the less-stable phase to a solid torus, which then shrinks and collapses into a ball. The mean curvature of a surface is given by $\kappa = 1/R_\perp + 1/R_\parallel$. For a solid torus with a large radius of the longitude curve $R_\parallel$, the curvature is dominated by that of the meridian curve with radius $R_\perp \ll R_\parallel$ and $\kappa \approx 1/R_\perp$. For the ball we have $R_\perp = R_\parallel$ so that, to satisfy the front velocity formula, the radius of the ball must be twice as large as the mean value of the radius of the disk-shaped cross-sectional area of the solid torus. In contrast to the unknot, the Hopf link is a stable pattern (Fig. 2). This is an interesting example of topological stabilization. The fact that the rings are linked leads to front repulsion at large ring diameters with small curvature. Note also that the density of the less-stable phase comprising the link is not uniform along the rings.

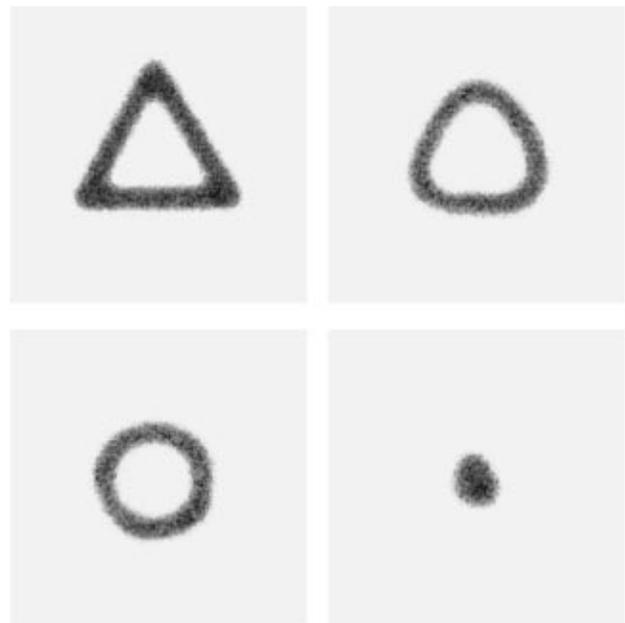

FIG. 1. Evolution of the unknot. The concentration of species $v$ is coded by gray shades, with black corresponding to high concentration and white to low concentration. The panels (left to right and top to bottom) correspond to times 1.0 (1000 steps), 53.0, 280.0, and 1000.0 ks. In the reduced units of the FHN equation, $1s = 0.029$. Mean-field parameter values are $\epsilon = 0.0137$, $a = 5.06$, and $b = 0.202$.

Knots of various types exist as stable structures and can be formed from initial conditions with the given topology. In Figs. 3 and 4 we show a stable trefoil and a figure-8 knot, respectively. In these figures we also display the knots as tubes built on curves $\mathbf{R}(\sigma) = [\mathbf{x}(\sigma), \mathbf{y}(\sigma), \mathbf{z}(\sigma)]$ spanning the pattern. An algorithm yielding $\mathbf{R}(\sigma)$ was constructed so that the parametrization of the line $\sigma$ is approximately proportional to the natural parametrization $s$.

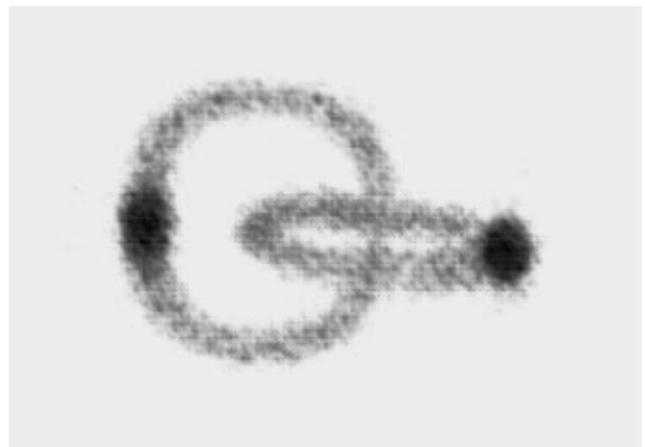

FIG. 2. A stable Hopf link. Parameter values same as Fig. 1, but the initial condition had the topology of a Hopf link composed of linear tubular segments.

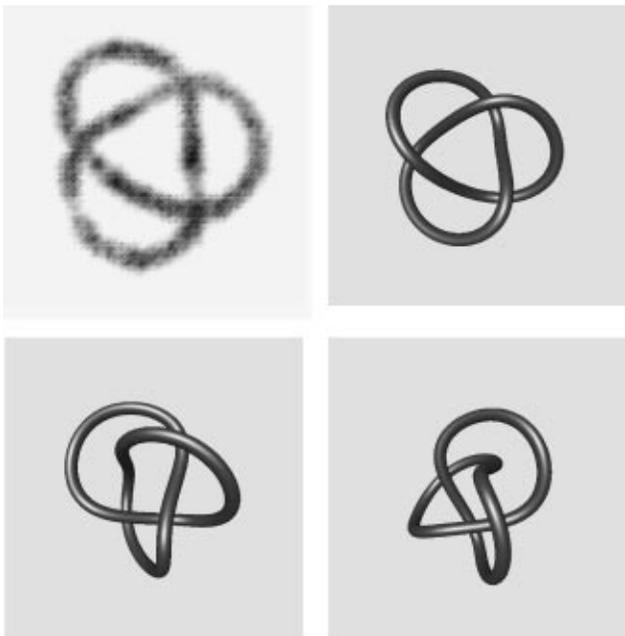

FIG. 3. A stable trefoil. The first panel (top left) displays the projection of $v$ concentration field as in Fig. 1. The second panel (top right) is a plot of the tubular neighborhood of the curve $\mathbf{R}(\sigma)$ extracted from an analysis of the concentration field. The bottom two panels show other projections of the trefoil. Simulations were carried out for the same parameter values as Fig. 1.

Insight into the geometrical structure of the knots can be obtained from the quantitative measures of a curve, the curvature and torsion [2,14],

$$\kappa = \frac{|\mathbf{R}_\sigma \times \mathbf{R}_{\sigma\sigma}|}{|\mathbf{R}_\sigma|^3}, \qquad \tau = \frac{(\mathbf{R}_\sigma \times \mathbf{R}_{\sigma\sigma}) \cdot \mathbf{R}_{\sigma\sigma\sigma}}{|\mathbf{R}_\sigma \times \mathbf{R}_{\sigma\sigma}|^2}.$$

The above expressions are invariant with respect to the change of variable $\sigma = \sigma(s)$ and thus are invariants of the curve. In Fig. 5 we plot the curvatures of the trefoil and the figure-8 knots. In both knots, $\kappa(\sigma)$ oscillates around approximately the same average value indicating that knots with varying complexity are composed of similar curved segments. In Fig. 6 we plot the torsion for these knots. A curve with a vanishing torsion lies in a plane [14] so that torsion may serve as a measure of nonplanarity. An analysis of the plot indicates that knots are composed of essentially planar segments connected to each other at points where torsion peaks.

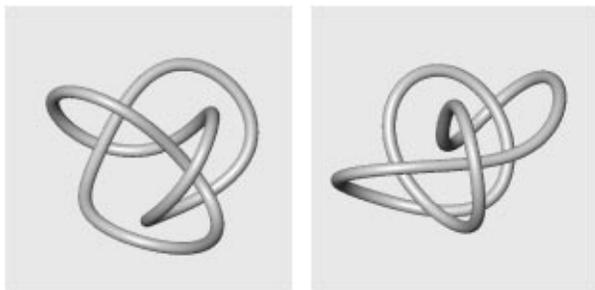

FIG. 4. Two projections of the figure-8 knot. Same parameter values as Fig. 1.

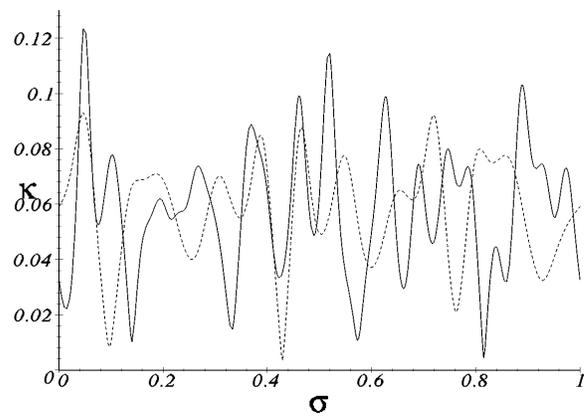

FIG. 5. Curvature along the trefoil (dashed line) and figure-8 (solid line) knots. Small values of the curvature correspond to segments of the knots encircled by a loop.

Because of the microscopic nature of the dynamics, in addition to the motion governed by the FHN mean field, there may exist noise-induced motion. The latter is the only mechanism for $\epsilon \gg 1$ when the pattern corresponds to a steady state at a local minimum of the free-energy functional. The motion should be consistent with the symmetry group of the object. For example, the symmetry group associated with rotations of the trefoil is $D_3$ [15], one axis of third order and three axes of second order [16]. For the Hopf link the symmetry group is $D_{2d}$, three orthogonal axes of second order and two planes of symmetry passing between two of them. There is no nonzero vector in $R^3$ which is invariant with respect to the action of these groups. From these arguments it follows that no motion of the trefoil or Hopf link is possible. The simplest knot with symmetry permitting motion is the figure-8 knot (see Fig. 4). While usual representations of the figure-8 knot possess $S_2$ symmetry, in our simulations the stable figure-8 knot has no symmetry. In either case, motion is possible but presumably it occurs on long time scales and was not observed in our simulations. This situation may be contrasted with the dynamics of singular filaments in excitable media that support propagating chemical waves [5]. The sign of rotation of the phase angle may be accounted for by the use of directed curves to represent the filaments. This lowers the symmetry, and motion due to the mean field is now possible and has been observed [5].

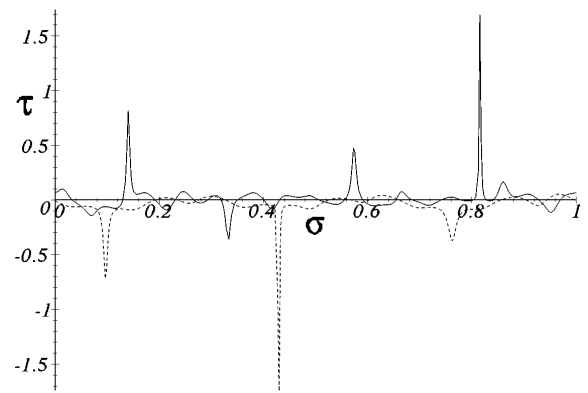

FIG. 6. Torsion along the trefoil (dashed line) and figure-8 (solid line) knots.

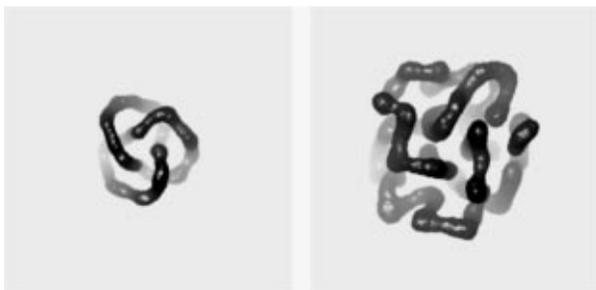

FIG. 7. An instability of the trefoil. The left and right panels refer to early and late times, respectively. Parameters are $\epsilon = 0.0137$, $a = 5.05$, $b = 0.195$.

Front instabilities in 3D lead to complex patterns. We have observed a variety of such instabilities by changing the system parameters. One such pattern is shown in Fig. 7 where the stable trefoil undergoes a fingering instability. New tubes stem from the initial trefoil while it itself becomes more and more wrinkled. In another instability, the tube forming the knot preserves its cross-sectional area but its length grows indefinitely. Similar types of instabilities have been seen for singular filaments in excitable media [5]. The linked and knotted patterns described here must form in bistable chemical media starting from random initial conditions, but it would be interesting to determine how experimental initial conditions might be chosen to guarantee evolution to knotted patterns.

This work was supported in part by a grant from the Natural Sciences and Engineering Research Council of Canada, an Ontario Graduate Scholarship (A. M.) and a Killam Research Fellowship (R. K.). We would like to thank S. G. Whittington, N. Margolus, and D. Gruner for useful discussions.